\documentclass[twocolumn,showpacs]{revtex4}

\usepackage{dcolumn}
\usepackage{amsmath}

\begin{document}

\preprint{CERN--TH/2001--055}

\title{A Note on Wandzura-Wilczek Relations}

\author{Patricia Ball}
\email{Patricia.Ball@cern.ch}
\affiliation{CERN/TH, CH--1211 Geneva 23, Switzerland \&\\ Institute for
  Particle Physics Phenomenology, University of Durham, Durham DH1~3LE, UK}
\author{Markus Lazar}
\email{lazar@itp.uni-leipzig.de}
\affiliation{Universit\"at Leipzig, Institut f\"ur
Theoretische Physik, Augustusplatz 10, D--04109 Leipzig, Germany}

\begin{abstract}
\noindent 
Wandzura-Wilczek (WW) relations between matrix-elements of bilocal
light-ray operators have
recently regained interest in connection with off-forward scattering
processes. Originally derived for matrix elements over leading-twist
operators, their generalization to off-forward and exclusive processes
gets complicated by the presence of higher-twist operators that are
total derivatives of leading-twist ones and do not contribute to
forward-scattering. We demonstrate that, for exclusive
matrix-elements, the inclusion of these operators into WW-relations is
essential for fulfilling constraints imposed by the conformal symmetry
of massless QCD. 
\end{abstract}

\pacs{11.30.Cp, 11.80.Et, 12.38.Aw, 13.60.Fz}

\maketitle

\noindent The success of QCD as fundamental theory of strong
interactions is intimately tied to its ability to describe hard
exclusive and 
inclusive reactions, which has been tested in numerous experiments. In
the corresponding kinematic regime, i.e.\ at large space-like virtualities,
the relevant amplitudes are dominated by singularities on the
light-cone, which, in the framework of a light-cone expansion
\cite{LCE}, can be
described in terms of contributions of definite twist. The notion of
twist, for local operators, 
has been introduced originally by Gross and Treiman \cite{GT} as
twist = dimension$\,-\,$spin;
it uses the irreducible representations of the orthochronous
Lorentz-group and, as such, is a Lorentz-invariant concept; its
generalization to non-local operators has been derived, in a
mathematical rigorous way, in \cite{GL1,GL2} for tensor operators up to
rank 2. 
This ``geometric'' twist relies solely
on geometry, i.e.\ the
properties of space-time, and is independent of the dynamics of any underlying
quantum field theory. An alternative approach to twist-counting in
hard reactions is
based on the light-cone quantization formalism \cite{LCQ}: quark
fields $\psi$ are decomposed into ``good'' and ``bad'' components, so
that $\psi = \psi_+ + \psi_-$, with $\psi_+ = \frac{1}{2}
\hat{p}\hat{z} \psi$ and $\psi_- = \frac{1}{2} \hat{z}\hat{p}\psi$
($\hat a = a^\mu\gamma_\mu$ for arbitrary 4-vectors $a$; $p$ and
$z$ are light-like vectors with $p\cdot z=1$; $z$ defines the
light-cone). As discussed in \cite{LCQ2}, a
 ``bad'' component $\psi_-$ introduces one unit of
twist; the physical interpretation of that ``dynamical'' twist is
that, in the infinite momentum frame,
it counts the powers of $1/Q$,
with which the corresponding matrix-elements appear in physical
scattering amplitudes; albeit being convenient, it is not a
Lorentz-invariant concept and does not
agree with the geometric twist. The mismatch between dynamical and
geometric twist becomes relevant once power-suppressed higher-twist
contributions are included and
leads to so-called Wandzura-Wilczek (WW) relations
between matrix-elements of operators of 
different dynamical, but identical geometric
twist, the prototype of which has been derived by Wandzura and 
Wilczek for the nucleon distribution  functions (DFs) $g_1$ and $g_2$ 
\cite{WW}. 
A systematic study of WW-relations in forward-scattering 
has been done in Ref.~\cite{gl01}.
The decomposition into operators of definite 
geometric twist has also been exploited by
Nachtmann to calculate exactly target-mass corrections to Bjorken-scaling in
deep-inelastic forward-scattering \cite{NM}. 
The successes of the purely geometric reasoning in applications to
forward-scattering matrix-elements has prompted several authors to
use it also for off-forward processes \cite{BRL} and
exclusive parton distribution amplitudes (DAs)\cite{MLrhoWF,MLWW}. 
The purpose of this letter is to argue that, despite its apparent success in
disentangling leading from higher-twist contributions to
forward-scattering matrix-elements, the generalization of WW-relations
to the non-forward and exclusive case requires the inclusion of
operators of higher twist. These operators arise from the dynamics of 
the underlying quantum field theory or, more precisely, the equations
of motion (EOM) and generate total translations. Their relevance for
preserving gauge-invariance in deeply-virtual Compton-scattering has
been discussed recently in \cite{gauge}; in this letter we demonstrate
that, for exclusive processes, the inclusion of these operators is
essential for fulfilling the constraints posed by
the dynamical symmetry of the theory, 
i.e.\ the invariance under collinear
conformal transformations in the case of massless QCD on the light-cone.

We center our discussion around the specific case of light
vector-meson DAs of dynamical twist-3, which are relevant for describing
light-cone dominated processes involving vector mesons like e.g.\ the
DIS-exclusive process $\gamma^* + N\to V+N$
and can be expressed in terms of matrix-elements of gauge-invariant
non-local operators sandwiched between the vacuum and the meson
state,
$
\langle 0 | \bar u(x) \Gamma[x,-x] d(-x) | \rho^-(P)\rangle,
$
where $\Gamma$ is a generic Dirac-matrix structure and $[x,y]$ denotes the
path-ordered gauge-factor along the straight line connecting the
points $x$ and $y$.
Specifying to chiral-even DAs, one can decompose
the relevant vector and axial-vector matrix-elements on the light-cone $z^2=0$
as~\cite{MLrhoWF,MLWW}
\begin{widetext}
$$
\begin{array}{l}
\displaystyle
\langle 0|\bar u(z) \gamma_{\mu} [z,-z]d(-z)|\rho^-(P,\lambda)\rangle
\ =\ f_\rho m_\rho\int_{-1}^1 d \xi
\bigg[ p_\mu\frac{e^{(\lambda)}\cdot z}{p\cdot z}
\hat\Phi^{(2)}(\xi) e_0(i\zeta\xi)\\[15pt]
\displaystyle+e^{(\lambda)}_{\perp\mu}
\Big\{\hat\Phi^{(2)}(\xi)e_1(i\zeta \xi)+\hat\Phi^{(3)}(\xi)
[e_0(i\zeta \xi)-e_1(i\zeta \xi)]\Big\}
-\frac{1}{2}z_\mu\frac{e^{(\lambda)}\cdot z}{(p\cdot z)^2} m_\rho^2
\left\{\hat\Phi^{(4)}(\xi)\Big[e_0(i\zeta \xi)-3e_1(i\zeta \xi)
+2\int_0^1\!\! d t\,  e_1(i\zeta \xi t)\Big]
\right.\\[15pt]
\multicolumn{1}{r}{\displaystyle-\hat\Phi^{(2)}(\xi)\Big[e_1(i\zeta \xi)
-2\int_0^1\!\! d t\, e_1(i\zeta \xi t)\Big]
+4\hat\Phi^{(3)}\Big[e_1(i\zeta \xi)
-\int_{0}^1\!\! d t\, e_1(i\zeta \xi t)\Big]\Big\}\bigg],}\\[15pt]
\displaystyle
\langle 0|\bar u(z) \gamma_{\mu}\gamma_5 [z,-z]d(-z)|\rho^-(P,\lambda)\rangle
\ =\ \frac{1}{2}\,f_\rho m_\rho
\epsilon_\mu^{\ \,\nu\alpha\beta}
e^{(\lambda)}_{\perp\nu} p_\alpha z_\beta
\int_{-1}^1 d \xi\, \hat\Xi^{(3)}(\xi) e_0(i\zeta \xi),
\end{array}
$$
where $\hat \Phi^{(d)}$ and $\hat \Xi^{(d)}$ contain only contributions from
geometric twist-$d$ and
 we use the following abbreviations:
\begin{align}
e^{(\lambda)}_{\perp \mu} = e^{(\lambda)}_{\mu} - p_\mu \,
\frac{e^{(\lambda)}\cdot z}{p \cdot z} ,\qquad \zeta=p\cdot z,
\qquad e_0(i\zeta \xi)=e^{i\zeta \xi},\qquad 
e_1(i\zeta \xi)=\int_0^1\!\!d t \,e^{i\zeta \xi t}.\nonumber
\end{align}
The same matrix-elements can also be expressed in terms of
contributions of definite dynamical twist as \cite{BBrhoWF,BBKT}
\begin{eqnarray}
\langle 0|\bar u(z) \gamma_{\mu} [z,-z]
d(-z)|\rho^-(P,\lambda)\rangle  
&=& f_{\rho} m_{\rho} \int_{-1}^{1} \!d\xi\, e^{i \xi p \cdot z} \left[ p_{\mu}
\frac{e^{(\lambda)}\cdot z}{p \cdot z}\,\hat\phi_{\parallel}(\xi) 
+ e^{(\lambda)}_{\perp \mu} \hat g_{\perp}^{(v)}(\xi)
- \frac{1}{2}z_{\mu}
\frac{e^{(\lambda)}\cdot z }{(p \cdot z)^{2}} m_{\rho}^{2}\,\hat
g_{3}(\xi) \right]\!,
\label{eq:vda}\\
\langle 0|\bar u(z) \gamma_{\mu} \gamma_{5}[z,-z]
d(-z)|\rho^-(P,\lambda)\rangle & = & 
\frac{1}{2}\, f_{\rho}
m_{\rho} \epsilon_{\mu}^{\phantom{\mu}\nu \alpha \beta}
e^{(\lambda)}_{\perp \nu} p_{\alpha} z_{\beta}
\int_{-1}^{1} \!d\xi\, e^{i \xi p \cdot z} \hat g^{(a)}_{\perp}(\xi);
\label{eq:avda}
\end{eqnarray}
\end{widetext}
$\hat\phi_\parallel$ has twist-2, $\hat g_{\perp}^{(v,a)}$ (dynamical) twist-3
and $\hat g_{3}$ has dynamical twist-4.
The relation between the two sets of DAs is
given by \cite{MLWW}
\begin{eqnarray}\label{eq:xyz}
\hat\phi_\|(\xi)&\equiv&\hat\Phi^{(2)}(\xi),\nonumber\\
\hat  g_\perp^{(v)}(\xi)&=&\hat\Phi^{(3)}(\xi) + \!\!\int\limits_\xi^{{\rm sign}(\xi)}\!\! \frac{d \omega}{\omega}
\Big(\hat\Phi^{(2)}-\hat\Phi^{(3)}\Big)\left(\omega\right),\nonumber\\
\hat g_3(\xi)&=&\hat\Phi^{(4)}(\xi) - \!\!\int\limits_\xi^{{\rm sign}(\xi)}\!\! \frac{d\omega}{\omega}\Big\{
\Big(\hat\Phi^{(2)}-4\hat\Phi^{(3)}+3\hat\Phi^{(4)}\Big)\left(\omega\right)
\nonumber\\
&&{}+ 2 \ln \Big(\frac{\xi}{\omega}\Big)
\Big(\hat\Phi^{(2)}-2\hat\Phi^{(3)}+\hat\Phi^{(4)}\Big)\left(\omega\right)
\Big\},\nonumber\\
\hat g_\perp^{(a)}(\xi)&\equiv&\hat\Xi^{(3)}(\xi).
\end{eqnarray}
Similar relations can be derived for chiral-odd DAs
 over the tensor and pseudoscalar currents. 

At this point we do observe a one-to-one correspondence between
the decomposition in terms of geometric twist DAs and DAs in dynamical
twist: there are four functions each. A difference does occur, 
however, as soon as we include
information on the dynamics of the theory. This information is
twofold, and is encoded in the dynamical symmetries of the theory on
the one hand and the equations of motion (EOM) on the other hand. As
for massless QCD at light-like distances, 
the relevant symmetry is the invariance under collinear conformal
transformations, i.e.\ the group SL(2,R)$\,\cong\,$SO(2,1), 
which is exact for the free theory and valid to
leading order in the perturbative expansion \cite{x}; for not too
small renormalization scales, the corresponding quantum number ``conformal
spin'', defined as 1/2$\,$(dimension + spin projection onto the line
$z_\mu$), is thus a good quantum number and allows a
partial-wave expansion of the corresponding amplitudes \cite{BBKT};
in the
limit $\alpha_s\to 0$ one obtains the so-called asymptotic DAs, which
are defined as the contribution with lowest conformal spin. Theoretical
calculations of the non-asymptotic corrections to the $\rho$-meson
DAs show that
they are small already at scales $\sim\,$1~GeV \cite{BBrhoWF}.
The EOM, on the other hand, allow one to establish
relations between e.g.\ bilinear operators of higher twist and
trilinear operators of leading twist 
and serve to identify the dynamically independent
degrees of freedom of a given DA \cite{BBKT}. In particular it turns out that
the basis of higher-twist DAs is
overcomplete: the number of
independent degrees of freedom is less than the number of independent
Lorentz-structures. This observation is of course not new; to the best
of our knowledge, the EOM have first been employed by Shuryak and
Vainshtein \cite{EOM} in relation with the 
WW-decomposition \cite{WW} of the (dynamical twist-3) spin-dependent nucleon
DF $g_2$, 
\begin{equation}
g_2(x_B) = -g_1(x_B) + \int_{x_B}^1 \frac{dy}{y}\, g_1(y)
+ \bar g_2(x_B),
\end{equation}
where $g_1$ is the leading-twist longitudinal spin DF and
$\bar g_2$ is the geometric twist-3 part. The original derivation of
Wandzura \& Wilczek  
involved only quark-quark operators, but Shuryak and Vainshtein noted
that, by virtue of
the EOM, the operators relevant for $\bar g_2$ can be written as
quark-quark-gluon operators whose matrix-elements are expected to be
small and thus can be neglected.

In general, however, the EOM do not only involve quark-quark-gluon
operators, but also quark-quark operators with total derivatives,
schematically
$
O^{(3)} = \tilde{O}^{(3),qqG} + \hat\partial^\mu O^{(2)}_\mu,
$
where $\tilde{O}^{(3),qqG}$ is an interaction-dependent operator and
  $O^{(2,3)}$ are quark-quark-correlation operators of geometric twist-2
  and 3, respectively. The important point is now that, although the
  above relation of course respects the geometric twist,
  matrix-elements can blur this decomposition, which
 happens whenever the matrix-elements over total
  derivatives do not vanish, i.e.\ for exclusive processes and
  off-forward scattering, 
  where the total derivative turns into a momentum (transfer).
To be specific, let us quote the
  formulas for the geometric twist-3 part of the vector and axial-vector
  currents~\cite{BBrhoWF,string}
\begin{widetext}
\begin{eqnarray}
 \Big[\bar u(-z)\gamma_\mu(\gamma_5) d(z)\Big]_{{\rm twist}\,3} &=&
{}-g_s\int_0^1 \!du\int_{-u}^u \!dv\,\bar u(-uz)
\Big[
u\tilde G_{\mu\nu}(vz)z^\nu\!\!\not\!z\gamma_5(\gamma_5)
 -ivG_{\mu\nu}(vz)z^\nu\!\!\not\!z(\gamma_5)
\Big]d(uz)
\nonumber\\
&&{}
+i\epsilon_{\mu}^{\phantom{\mu}\nu\alpha\beta}\int_0^1 udu\,
z_\nu\hat\partial_\alpha
\Big[\bar u(-uz)\gamma_\beta\gamma_5(\gamma_5) d(uz)\Big]\,,
\label{twist3}
\end{eqnarray}
\end{widetext}
where $G_{\mu\nu}$ is the gluonic field strength, $\tilde G_{\mu\nu}
=(1/2)\epsilon_{\mu\nu\alpha\beta}G^{\alpha\beta}$ its dual, and
$\hat\partial_\alpha$ is the derivative over the total translation:
$$
\hat\partial_\alpha\Big[\bar u(-z)\gamma_\beta d(z)\Big] \equiv
\left.\frac{\partial}{\partial y_\alpha}
\Big[\bar u(-z+y)\gamma_\beta d(z+y)\Big]\right|_{y\to 0}.
$$
We would like to stress that, 
from the group-theoretical point of view, Eq.$\,$(\ref{twist3}) is 
a genuine geometric twist-3 relation.
Taking matrix-elements and neglecting
 the manifestly interaction-dependent quark-quark-gluon
 operators, whose numerical contribution is small \cite{BBKT}, 
one finds the ``WW-type'' relations \cite{BBrhoWF}
\begin{eqnarray}
   \hat g_\perp^{(v),{{\rm dWW}}}(\xi) &=&
 \int\limits_{-1}^{\xi} d\omega\, \frac{\hat\phi_\parallel(\omega)}{1-\omega}
 + \int\limits_{\xi}^{1} d\omega\, \frac{\hat\phi_\parallel(\omega)}{1+\omega},
  \nonumber\\
   \hat  g_\perp^{(a),{{\rm dWW}}}(\xi) &=&
   \int\limits_{-1}^{\xi}\!\! d\omega\,
     \frac{1-\xi}{1-
\omega}\,\hat\phi_\parallel(\omega)+\int\limits_{\xi}^{1} \!\!
d\omega\, \frac{1+\xi}{1+\omega}\,\hat\phi_\parallel(\omega).\nonumber\\[-10pt]
  \label{WW1}
  \end{eqnarray}
A comparison with (\ref{eq:xyz}) reveals that $\hat g_\perp^{(a)}$,
although manifestly of geometric twist-3, has ``inherited'' twist-2
contributions from the total derivative; also $\hat g_\perp^{(v)}$ contains
such terms. The important point to note is that
the above WW-relations are consistent with the conformal expansion in
the sense that (a) inserting the asymptotic DA 
$\hat\phi_\parallel^{\rm as} = 3(1-\xi^2)/4$ yields the asymptotic DAs
\begin{equation}
g_\perp^{(v),{\rm as}} =  \frac{3}{4}\, \left( 1 + \xi^2 \right),\quad
g_\perp^{(a),{\rm as}} =  \frac{3}{4}\,(1-\xi^2),
\end{equation}
as required by conformal expansion,
cf.$\,$Ref.$\,$\cite{BBKT}, and that (b)
contributions of higher conformal spin to $\hat\phi_\parallel$ translate
into contributions of the same conformal spin to
$\hat g_\perp^{(v,a)}$.
The interaction-dependent operators only
contribute at non-leading conformal spin.

The above relations include the contribution of the twist-3
total-derivative operator in (\ref{twist3}).
What happens if, in the original spirit of WW, one only
includes geometric twist-2 operators in the WW-relations? As shown in
\cite{MLWW}, one finds
\begin{eqnarray}
\hat g_\perp^{(v),{{\rm gWW}}}(\xi) &=& \int_\xi^{\rm{sign}(\xi)}
\frac{d\omega}{\omega}\, \hat{\phi}_\parallel(\omega),\nonumber\\
\hat g_\perp^{(a),{{\rm gWW}}}(\xi) &=& 0,
\end{eqnarray}
which, inserting $\hat{\phi}_\parallel^{\rm as}$, yields
\begin{equation}
\hat g_\perp^{(v),{{\rm gWW}}}(\xi) = \frac{3}{8}\,\left(
  \xi^2 -1 - 2 \ln\,\frac{\xi}{\rm{sign}(\xi)}\right),
\end{equation}
which exhibits a logarithmic singularity at $\xi=0$.

These results are quite different from those obtained from 
conformal expansion. The argument of WW for neglecting the twist-3
operators, corroborated by Shuryak and Vainshtein, was that they
are equivalent to  quark-quark-gluon operators whose matrix-elements  can
be neglected numerically. This
argument, however, does no longer hold in exclusive kinematics, where
twist-3 operators with 
total derivatives induce contributions that are as large as those from
twist-2. 
Thus, the contributions
$\hat g_\perp^{(v),{\rm tw3}}(\xi) 
=\hat g_\perp^{(v)}(\xi)- \hat g_\perp^{(v),{{\rm gWW}}}(\xi) $
and $\hat g_\perp^{(a),{\rm tw3}}(\xi) =\hat g_\perp^{(a)}(\xi)$
are not small numerically.

In addition we have demonstrated above that the geometric
WW-relations violate the restrictions imposed by conformal symmetry
and yield (artificial) singularities that are cancelled {\em exactly} 
by total-derivative operators of geometric twist-3. 
We conclude that analyses based on geometric twist-2 are of rather limited 
use in deriving WW-relations between DAs of different dynamical twist and 
that it is essential to include twist-3 operators containing total 
derivatives 
in order to preserve the symmetries of the theory. This result is
complementary to that obtained in Ref.$\,$\cite{gauge} that twist-3
total-derivative operators are needed to restore gauge-invariance of
physical amplitudes in off-forward kinematics.

Let us finally also comment on the possible extension of 
WW-relations to twist-4. At this order in the twist-expansion,
trace-subtractions of leading twist-2 operators become relevant and
give rise to two different types of relations: one gives the geometric
twist-2 part of the dynamical twist-4 DA $\hat g_3$ and was obtained
in \cite{MLWW}:
\begin{equation}\label{star}
\hat g_3^{\rm tw2}(\xi) = -\int_{\xi}^{{\rm sign}(\xi)}
\frac{d\omega}{\omega} \, \left\{ 1 + 2 \ln \left(
    \frac{\xi}{\omega}\right)\right\}\hat\phi_\parallel(\omega).
\end{equation}
This expression shows again a logarithmic singularity that is not
present in the full expression for $\hat g_3$ obtained in \cite{BBT4}
using the EOM. Trace-subtractions of the twist-2 operator also give
rise to so-called ``kinematical'' target-mass
corrections. For the
forward-scattering case, contributions of this type have been
considered by Nachtmann.
For the exclusive case,
the corresponding operators have been considered in \cite{excl,tocome}, and
for off-forward scattering, the  resummation has been
done in \cite{BM}. The relevance of such a procedure remains,
however, unclear. For, in addition to the geometric
mass-corrections, one also has also ``dynamic'' mass-corrections 
from operators $\sim x^{2n}\partial^{2n} O^{(2)}$
which are of geometric twist-$(2n+2)$ and such that are ``hidden'' in
the twist-4 quark-quark-gluon operators entering by the EOM. This is
the exact analogue of what happens at twist-3: decomposition of the
relevant operators in terms of irreducible representations of the
Lorentz-group gives only part of the information: the EOM have to be
applied in order to obtain gauge- and conformal-invariant
results. Indeed, the authors of \cite{BM} observe that their results for
mass-corrections violate gauge-invariance. Dynamical
mass-corrections have so far only been considered for the exclusive
case, Ref.~\cite{BBT4}. Again, it was not possible to formulate the
twist-4 analogue of Eq.$\,$(\ref{twist3}), with a clean separation of
interaction-dependent and total-derivative terms; instead, one had to
rely on a cumbersome local expansion that was used to obtain results
for the next-to-leading order in the conformal expansion.
Numerically, these corrections turned out to be relevant. We conclude
that, at least for exclusive vector-meson DAs, a resummation of
mass-corrections induced by trace-subtractions in the leading twist
matrix element and the higher-twist operators containing total derivatives
are relevant for a good approximation.

\begin{acknowledgments}
\noindent P.B.\ is supported by DFG through a Heisenberg fellowship.
M.L.\ acknowledges the Graduate College
``Quantum field theory'' 
of Leipzig University for financial support and the CERN/TH-division 
for  support during his stay in Geneva.
\end{acknowledgments}

\end{document}